\documentclass[fleqn,12pt,twoside, amsmath,amssymb,psfig]{article}
\usepackage[headings]{espcrc1}
\usepackage{graphicx}
\usepackage{bm}

\readRCS
$Id: espcrc1.tex,v 1.2 2004/02/24 11:22:11 spepping Exp $
\newcommand{\dif}{\mathrm{d}}
\newcommand{\I}{\mathbf{i}}
\newcommand{\J}{\mathbf{j}}
\newcommand{\zero}{\mathbf{0}}
\newcommand{\m}{\mathbf{m}}

\newcommand{\R}{\mathbf{r}}

\newcommand{\nI}{n_{\I}}
\newcommand{\nJ}{n_{\J}}
\newcommand{\sI}{s_{\I}}
\newcommand{\sJ}{s_{\J}}
\newcommand{\qI}{q_{\I}}
\newcommand{\qJ}{q_{\J}}
\newcommand{\Hn}{{\mathcal H}_{\mathrm{n}}}
\newcommand{\HC}{{\mathcal H}_{\mathrm{C}}}
\newcommand{\betan}{\beta_{\mathrm{n}}}

\newcommand{\betaC}{\beta_{\mathrm{C}}}
\newcommand{\En}{E_{\mathrm{n}}}
\newcommand{\EC}{E_{\mathrm{C}}}
\newcommand{\HnC}{{\mathcal H}_{\mathrm{n+C}}}
\newcommand{\HFe}{{\mathcal H}_{\mathrm{Fe}}}

\newcommand{\M}{{\mathcal M}}

\newcommand{\sumNIJ}{\sum^V_{\I,\J}\hspace{-2pt}' }
\newcommand{\SUMNIdifJ}{\sum^V_{\I\neq\J}}
\newcommand{\SUMNIJ}{\sum^V_{\I,\J}}

\newcommand{\AmS}{{\protect\the\textfont2
  A\kern-.1667em\lower.5ex\hbox{M}\kern-.125emS}}

\runtitle{	Thermodynamics of compact-star matter within
		an Ising approach}
\title{		Thermodynamics of compact-star matter within
		an Ising approach}

\runauthor{	P. Chomaz}
\author{	P. Chomaz\address[MCSD]{Ganil(DSM-CEA/IN2P3-CNRS), Blvd. H. Becquerel, BP 55027, F-14076 Caen c\'edex5, France},
        	C. Ducoin\addressmark[MCSD]\address[2]{LPC(IN2P3-CNRS/Ensicaen et Universit\'e), F-14076 Caen c\'edex, France},
        	F. Gulminelli\addressmark[2]\footnote{member of the Institut Universitaire de France},
        	K. Hasnaoui\addressmark[MCSD]
        and
 		P. Napolitani\addressmark[MCSD]\addressmark[2]
 	}

\begin{document}

\maketitle

\begin{abstract}
	In the formation and evolution of compact stars,
nuclear matter explores high thermal excursions and is the site
of intense neutrino emission.
 	Neutrino transport as well as structural properties of
this matter depend on the presence of inhomogeneous phases (named 
``pasta'' phases), which are the result of Coulomb frustration of the 
Liquid-Gas phase transition.
	We take into account charge fluctuations by employing 
a frustrated lattice-gas model to which we impose a neutrality constraint 
by the addition of an homogeneous
background of charge, representing delocalised electrons.
	Within this schematic model we highlight a generaic 
feature of the phase-transition phenomenology: 
the temperature interval where pasta phases are formed is 
enhanced by Coulomb-frustration effects. This result is at variance 
with the behaviour of frustrated ferromagnetic systems as well as 
hot nuclei and mean-field approaches.
	Moreover, the region of phase coexistence is not found to 
end upon a critical point, indicating that no critical opalescence
can occur in compact-star matter.
	
\end{abstract}
%
%
\section{Introduction: stars and atomic nuclei}

	Nuclear matter present in compact and hot astrophysical 
objects generated in a gravitational collapse suffers from high
thermal excursions.
In the accessible range of temperature and densities, nuclear matter
presents phase transitions characterized by density fluctuations.
	Because of the presence of the long range 
	Coulomb interaction, those charge fluctuations give rise to so 
called ``frustration effects''~\cite{Negele1973}.
	Under these conditions, the system organises in 
inhomogeneous phases, named ``pasta phases'' due to their unusual
topologies
~\cite{pasta}.
	The evolution of these structures with density is 
expected to be connected to the properties of 
the inner neutron-star crust.
The survival of pasta-phases at finite temperature has been inferred from recent 
molecular-dynamics simulations~\cite{mol_din}.
	As these complex structures are related to 
coherent neutrino scattering~\cite{Horowitz2004},
their evolution with density and temperature or, more generally, the 
underlying equation of state, becomes a key quest for exploring 
several interconnected astrophysical phenomena, 
like the dynamics of supernova explosion 
and the thermodynamics of proto-neutron-stars cooling.

	A tempting connection can be searched with the physics of 
atomic nuclei, which are experimentally accessible.
	Comparable conditions of temperature and density can 
be probed in rather violent processes like nuclear 
multifragmentation.
	However, compact-star matter differs from normal nuclear matter 
	due to the presence of electrons.
	They constitute an incompressible degenerated gas, 
which establishes a condition of global charge neutrality at large 
scale.


	To understand the phenomenology of phase transitions
	in compact stars, we consider an Ising model. Indeed widely used mean-field 
	approximations ~\cite{Lattimer1985,Haensel2000,Ducoin2006}
	are known to fail to describe critical phenomena in 3D.
	The Ising model is specifically suited for the analysis of charge 
fluctuations by exact calculations, since phase transitions are universal 
phenomena. Moreover, it was largely employed in the 
study of ferromagnetic systems subjected to 
frustration~\cite{Grousson} as well as for nuclear 
matter~\cite{Gulminelli2003} where, in the form of the lattice-gas 
model, magnetic spins are replaced by site occupations.
%
%
\section{Ising analogue to compact-star matter}

	We construct a cubic periodic lattice composed of $V$ 
sites, each characterised by a position $\I$ and an occupation 
number $\nI=1$ or $0$, to indicate the presence or the absence of 
positive charge, respectively.
	This distribution of positive charge simulates the 
distribution of nuclear-matter charge (neutron and protons without 
distinction are considered as nucleons with an effective charge Z/A).
	The strongly incompressible gas of electrons is represented 
by a uniform distribution of negative charge~\cite{Maruyama2005} 
 imposing a strict condition of neutrality.
	In any site we find therefore an effective charge 
$\qI = \nI - \bar{n} $, where $\bar{n} =\sum^{V}_{\J}\nJ/V$
is the charge per site of the background of negative charge.

	The schematic Hamiltonian $\HnC=\Hn+\HC$ with
\begin{equation}
	\Hn =
		\frac{\epsilon}{2}
		\sumNIJ \nI\nJ
	,\quad
	\HC =
		\frac{\lambda\epsilon}{2}
		\SUMNIdifJ \frac{\qI\qJ}{r_{\I\J}}
\label{eq:Hamiltonian}
\end{equation}
is introduced to study 
the interplay of nuclear-like ($\Hn$) and Coulomb-like forces ($\HC$).
	$\sum'^V_{\I,\J}$ is a sum extended over closest
neighbors, and $r_{\I\J}$ is the distance between sites $\I$ and $\J$.
	The short-range and long-range interactions are characterized
by the coupling constants $\epsilon$
and $\lambda\epsilon = \alpha\hbar c \rho_0^{1/3} x^2$ respectively, where $\rho_0$
is the nuclear saturation density, and $x$ is the proton fraction.
	The ratio of the two coupling constants $\lambda$ measures the 
strength of frustration.
	In the sequel, $x$ is fixed to $1/3$, a value expected for
proto-neutron stars.
	$\HC$ can be rewritten in terms of
occupation products times the geometric constants $C_{\I\J}$ as
\begin{equation}
	\HC =
		\frac{\lambda\epsilon}{2}
		\SUMNIJ \nI\nJ C_{\I\J}
	.
\label{eq:HC}
\end{equation}
	Translational invariance, which is ensured
by imposing the distance $\R_{\I\J}$ to be the shortest 
between $\I$ and $\J$ in the periodic space, imposes 
$\sum^{V}_{\I} C_{\I\J}=0$.
	Hence, $C_{\I\J}$ can be calculated by adopting periodic 
boundary conditions in the form
\begin{equation}
		C_{\I\J} = D_{\I\J} - {\mathcal D}
	,\quad\mathrm{ with }\quad
		{\mathcal D}=\frac{1}{V}\sum_{i'}^{V}D_{\zero\I'}
	,
\label{eq:Cij}
\end{equation}
	where $D_{\I\J}=0$ if $\I=\J$, and 
$D_{\I\J}=|\R_{\I\J}|^{-1}$ otherwise; ${\mathcal D}$ is calculated
with respect to any position $\zero$. 
%

	$\HC$ preserves all the characteristic symmetries of the 
Ising model and the Coulomb energy is therefore invariant for 
lattice-occupation reversal.
	We should also point out that, despite some tempting 
similarities, the correspondent of eq.(\ref{eq:Hamiltonian}) in spin 
variable $\sI=\nI-1/2$ and $\rho=\M/V+1/2$ with the 
magnetization $\M=\sum_{\I}^{V}\sI$  
is not identical to the Hamiltonian of the 
frustrated Ising ferromagnet~\cite{Grousson}, defined
by
\begin{equation}
	\HFe =
		\frac{\epsilon}{2}
		\sumNIJ \sI\sJ
		+
		\frac{q\epsilon}{2}
		\SUMNIJ \sI\sJ D_{\I\J}
	.
\label{eq:Fe}
\end{equation}
	Indeed, 
	$\HFe$ 
	and $\HnC$ are related by
\begin{equation}
    \HFe=
        \HnC
        + \frac{\lambda\epsilon}{2}{\mathcal D} \M^2
        + 3\epsilon\M
        - \frac{3\epsilon}{4}V
    ,
\label{eq:isomorphism}
\end{equation}
This connection includes a term in $\M^2$, scaling with $V^2$. Such term
deeply modifies the thermodynamics in a finite system, and imposes
 $\M=0$ at the thermodynamic limit for $\HFe$.
	The strict constraint $\M=0$ on the order parameter $\M$ 
	leads to a substantial
modification of the thermodynamics of the system\cite{Carmona2003}.
The phase diagram of the frustrated ferromagnet~\cite{Grousson}, where
, as in nuclei, the Coulomb interaction
reduces the phase-coexistence region (i.e. decreases the limiting 
temperature), cannot be directly exported to our system $\HnC$.
 	
%
%
\section{Simulation in the multi-grandcanonical ensemble}
%
%
\begin{figure}[pb!]
\begin{center}
\vspace{-0.3 cm}
\includegraphics[angle=0, width=0.35\textwidth]{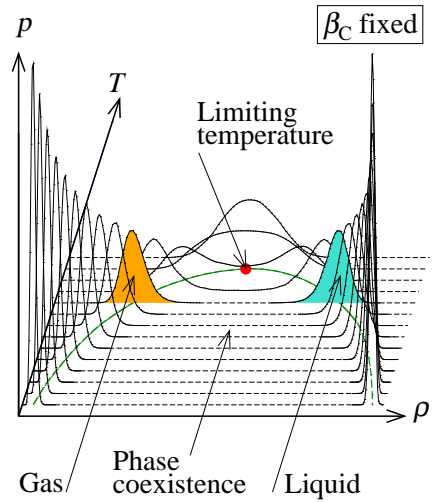}
\end{center}
\vspace{-0.7 cm}
\caption{\label{fig:fig1}
	Example of a Metropolis calculation ($\mu=\mu_c$ ).
	Density distribution for a fixed value of $\betaC$ and a 
	series of temperatures, each one associated to one 
	spectrum.
	The ensemble of explored temperatures and densities defines
	the phase diagram.
	The ridge of the distribution defines three branches 
	joining at the limiting temperature.
	The phase-coexistence region is delimited by the gas (low 
	density) and liquid (high density) branches.
}
\end{figure}

	For the numerical simulation of the $\HnC$ system, in order
to take into account the long-range Coulomb 
interactions in a rapidly converging form, the finite lattice is 
repeated in all three directions of space a large number $R$ of 
times.
    Each site $\I$ has $R$ replicas of itself, each one displaced 
from $\I$ of a vector $\m L$, where $\m$ has integer components and 
$L = \sqrt[3]{V}$.

	To treat a frustrated system, described by the
interplay of several interactions, the multi-(grand) canonical 
ensemble\cite{Gulminelli2003} is specifically suited.
	It consists in associating a Lagrange multiplier to each 
component of the Hamiltonian that we want to treat as a separate 
observable.
	In particular, it makes possible to construct one single 
phase diagram for both neutral and charged matter by associating 
one Lagrange multiplier, $\betan$, to the nuclear energy (related
to $\Hn$) and another, $\betaC$, to the Coulomb energy $\EC$
(related to $\HC$).
	The multi-canonical partition sum reads:
\begin{equation}
	Z_{\betan,\betaC,V}(\rho V,V) =\int W_V(\En,\EC,\rho V,V)
	e^{-\betan \En -\betaC \EC } \dif\En\dif\EC   \nonumber\, 
	,
	\label{cano}
\end{equation}
where $W_V(\En,\EC,\rho V,V)$ is the density of states.
	For any finite value of $\betaC$, we treat the system as 
equilibrated at the temperature $T=1/\betan$ and described by an 
equivalent effective charge 
$q_{\mathrm{eff}}^2 = \lambda\betaC/\betan$.
	When $\betaC = 0$, the Coulomb energy $\EC$ does no more 
influence the partition sum and the system behaves as uncharged,
reducing to the standard Ising model. 
	The generalised grand potential is defined as
\begin{equation}
	Z^G_{\betan,\betaC,\alpha}(V)=\int Z_{\betan,\betaC}(\rho V,V)
	e^{-\alpha \rho V}\dif\rho \, 
	, 
	\label{macro}
\end{equation}
where the fugacity is linked to the chemical potential by 
$\alpha = -\betan \mu$.
	When $\betaC = \betan$ the ensemble coincides with the
conventional grand-canonical form.

	We sampled the density distribution from eq.(\ref{macro}) 
for a chemical potential value $\mu=\mu_c=3\epsilon$, 
by employing a standard Metropolis technique~\cite{Chomaz2002}.
	An example is presented in Fig.~\ref{fig:fig1} where, for
a fixed value of $\betaC$, several density distributions are
calculated for different temperatures $T = 1/\betan$.
	All bimodal distributions present a peak for the gas phase
(low-density) and one for the liquid (high-density) phase.
	Their height is equal because of the choice of $\mu=\mu_c$.
	All distributions are composed together to draw a phase 
diagram: the ridge of the overall distribution of density and 
temperature delimits with its gas branch and liquid branch the 
phase coexistence.
	The two branches join at the limiting temperature.
%

\section{Results}

\begin{figure}[bp!]
\begin{center}
\vspace{-0.3 cm}
\includegraphics[angle=0, width=0.65\columnwidth]{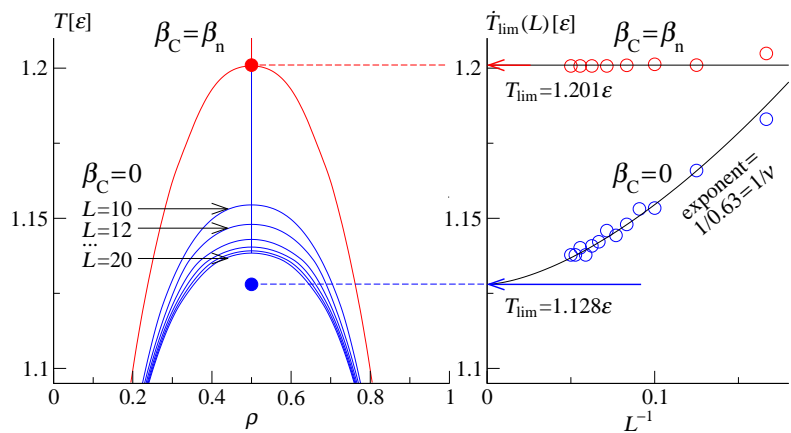}
\end{center}
\vspace{-0.7 cm}
\caption{\label{fig:fig2}
	\textbf{Left panel}.
	Calculations of phase diagrams for the systems
 	$\betan=0$ (Ising-like) and $\betan=\betaC$ (frustrated).
	Calculations for a series of 
	different lattice sizes $L$ are associated to different 
	limiting temperatures for the system $\betan=0$; 
	they overlap for the system $\betan=\betaC$.
	\textbf{Right panel}.
	For the system $\betan=0$ the test of the 
	finite-size scaling law
	$\dot{T}_{\mathrm{lim}}(L) - T_{\mathrm{lim}} \propto L^{-1/\nu}$
	gives Ising values for the critical exponent $\nu$ and the 
	critical temperature $T_{\mathrm{lim}}$.
	No scaling is observed for the system $\betan=\betaC$: this
	is a first indication of the quench of criticality in 
	presence of a Coulomb field.
}
\end{figure}

%
%
%
	For two cases, either $\betan=0$ (corresponding to 
the standard Ising model), or $\betan=\betaC$, and for a series of 
different lattice sizes $L$, we collected a series of calculations
of the phase diagram.
	The results are schematized in Fig.~\ref{fig:fig2}.
	The phase-coexistence region in the system $\betan=0$
shrinks for progressively larger lattice sizes according to the
characteristic finite-size scaling behaviour of the Ising system.
	In particular, since the asymptotic value of the limiting 
temperature 
$T_{\mathrm{lim}}=\lim_{L\rightarrow\infty} \dot{T}_{\mathrm{lim}}(L)$ 
corresponds to a critical point, finite-size scaling requires that 
$\dot{T}_{\mathrm{lim}}(L)$ evolves as 
$\dot{T}_{\mathrm{lim}}(L) - T_{\mathrm{lim}} \propto L^{-1/\nu}$~\cite{scaling}.
	This is comfirmed by the calculated values of the critical 
exponent $\nu$ and the critical point ${T}_{\mathrm{lim}}$,
which correspond to the Ising values.
%
%

	At variance with the Ising behaviour, the phase diagram of
the system $\betan=\betaC$ does however not vary with the lattice 
size.
	This first sign of incompatibility of the frustrated system 
with the Ising model, is also a first indication of quenching of 
the critical behaviour.
	$\nu$ rules in fact the divergence of the correlation 
length at the critical point, which evolves as 
$\xi\propto t^{-\nu}$ with $t = T/T_{\mathrm{lim}}-1$.
	If a critical point existed, 
it would correspond to a very large value for $\nu$, which
would then be hardly compatible with a diverging correlation length. 

	The major result we infer from comparing the two systems 
and the corresponding asymptotic limiting temperature, 
is that the coexistence region expands under the action of a 
Coulomb field.
%
%
%
\begin{figure}[pb!]
\begin{center}
\vspace{-0.3 cm}
\includegraphics[angle=0, width=0.85\columnwidth]{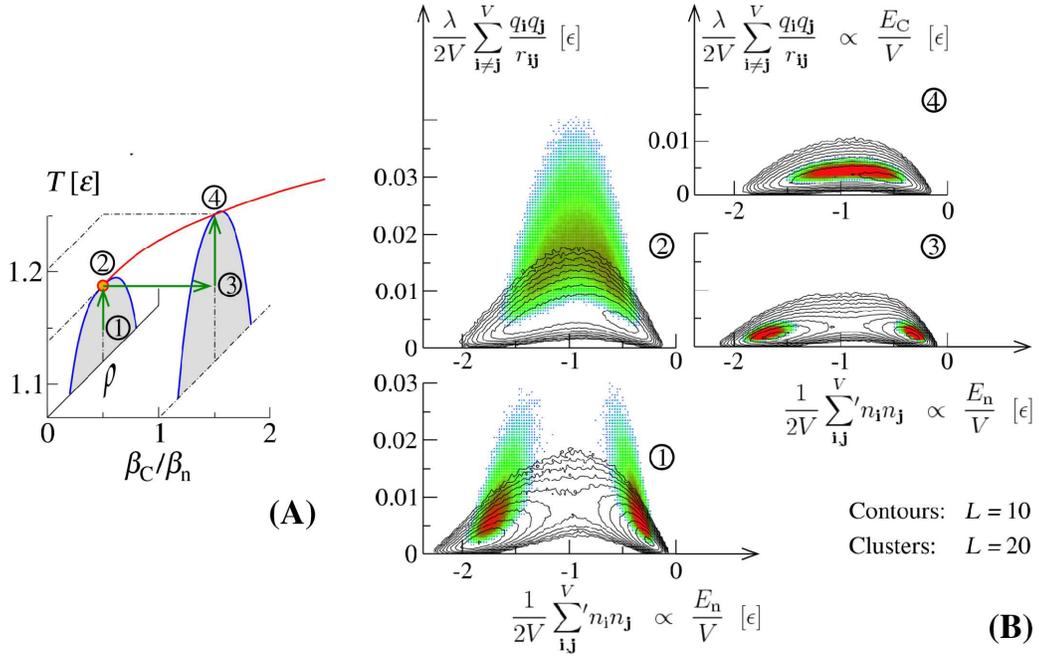}
\end{center}
\vspace{-0.7 cm}
\caption{\label{fig:fig3}
	\textbf{(A)}
	Increase of the limiting point and expansion of the 
	phase-coexistence region with increasing frustration (i.e.
	for higher $\betaC/\betan$ ratios). Four points determine
	different thermodynamical situations: 
	points (1) and (3) belong to the coexistence region,
	points (2) and (4) are at the limiting temperature;
	while points (1) and (2) refer to the system $\betaC=0$ 
	(Ising-like), points (3) and (4) refer to the system 
	$\betaC=\betan$ (frustrated).
	The calculation refers to $L=10$;
	\textbf{(B)}
	In correspondence with the points marked in panel (A),	
	the probability distribution is evaluated for the two
	Hamiltonian components, $(1/2V)\sumNIJ \nI\nJ$, which
	corresponds to the nuclear-energy density $\En/V$, and 
	$(\lambda/2V)\SUMNIdifJ \qI\qJ/r_{\I\J}$,
	representing the Coulomb-energy density $\EC/V$ 
	in the case of the frustrated system.
	Contour plots refer to $L=10$;
	logarithmic cluster plots refer to $L=20$.
	Points (1) and (3) manifest bimodal patterns.
	The critical scaling effect exhibited at point (2)
	quenches at point (4).
}
\end{figure}

	The connection between the increase of the limiting point	
and charge fluctuations can be searched in the evolution of the
event distribution with temperature and Coulomb-field strength.
	Fig.~\ref{fig:fig3}A illustrates the evolution of the 
limiting temperature with the strength of the Coulomb field and 
the phase coexistence is indicated for the uncharged system 
$\betaC=0$ and the frustrated system $\betaC=\betan$, in 
correspondence with the phase diagrams shown in 
Fig.~\ref{fig:fig2}.
	We analyzed the event distribution along the path 1-2-3-4, 
determined by four different thermodynamic situations.

	In Fig.~\ref{fig:fig3}B the distributions are shown, 
as a function of the two Hamiltonian components, 
$(1/2V)\sumNIJ \nI\nJ$ and 
$(\lambda/2V)\SUMNIdifJ \qI\qJ/r_{\I\J}$.
	Point (1) belongs to the region of phase coexistence 
sampled for the uncharged system $\betaC=0$ and presents a bimodal
pattern with respect to the abscissa, which corresponds to the 
nuclear-energy density $\En/V$, so that pure liquid or gas 
partitions are preferred to mixed events.
	Point (2) corresponds to the limiting temperature of 
the uncharged system $\betaC=0$, where mixed partitions are 
favoured.
	To pass from point (2) to point (3) the temperature
is left unchanged while $\betaC$ is increased from $0$ till it
equals $\betan$.
	Once introduced the Coulomb field, 
$(\lambda/2V)\SUMNIdifJ \qI\qJ/r_{\I\J}$ represents the 
Coulomb-energy density $\EC/V$, which should be minimized in the
frustrated system.
	In order to satisfy this constraint, the positive-charge 
distribution searches uniform patterns, which are better 
compensated by the uniform background of negative charges.
	As a result, the system moves to the phase-coexistence region to search
pure partitions, and bimodality appears.
	In the frustrated system, the temperature should be 
further increased with respect to the limiting temperature of the 
system $\betaC=0$, point (3), in order to find the limiting 
temperature, as indicated at point (4).

	Fig.~\ref{fig:fig3}B presents the event distributions 
sampled for two different lattice sizes, in order to test
scaling effects which could indicate the presence of critical
points.
	At a critical point the correlation length $\xi$ diverges
and determines the exponential decay of the correlation 
function $\sigma(\R_{\I,\J}) = 
\langle \nI\nJ \rangle - \langle \nI \rangle \langle \nJ \rangle$
according to the expression  
$\sigma(r) \propto   e^{-r/\xi} \cdot r^{-(D-2+\eta)}$,
where $D$ is the space dimension and $\eta$ is a critical exponent.
	On the basis of the properties of the $C_{\I\J}$ matrix, 
$\sigma(r)$ is related to the mean Coulomb-energy density by 
\begin{equation}
	\Bigg\langle \frac{\lambda\epsilon}{2V}
	\SUMNIdifJ \frac{\qI\qJ}{r_{\I\J}} \Bigg\rangle =
	\Bigg\langle\frac{\EC}{V}\Bigg\rangle =
	\frac{\lambda\epsilon}{2} \left(\sigma(0){\mathcal D} + 
	\sum_{\J \neq \zero}^{V}
	\frac{\sigma(\R_{\zero\J})}{r_{\zero\J}}\right)
        .
\label{eq:energydens}
\end{equation}
	It is therefore typical of a critical point to exhibit
a divergence of the quantity 
$\langle \SUMNIdifJ \qI\qJ/r_{\I\J} \rangle$ for progressively 
larger lattice sizes.
	This indicates that, at point (2), the limiting point is
compatible with a critical point, as we already expected from the
finite-size scaling law tested in Fig.~\ref{fig:fig2} for the
critical exponent $\nu$.
	On the contrary, no scaling is observed at the point (4).
	Indeed, as indicated by Eq.(\ref{eq:energydens}), when the 
Coulomb field is introduced in the system, the diverging quantity 
would be the Coulomb-energy density $\EC/V$.
	To avoid such a singularity, the correlation length keeps
a finite value at the limiting temperature and the critical 
character of the limiting temperature  is lost~\cite{Chomaz2005}.
	A more formal discussion on the suppression of criticality
on the basis of finite-size-scaling analysis is presented in 
ref.~\cite{in-progress}.

%
%
\section{Conclusions}

	By employing a specific model for studying the 
thermodynamical features of a frustrated system, we could define
general properties of a neutral system in presence of charge 
fluctuations, in analogy to compact-star matter.
	The study of the phase diagram and the phase-transition 
phenomenology indicated that the introduction of a Coulomb field 
has the effect of increasing the limiting temperature without
preserving any critical character.

	This result is the opposite as found for many other 
physical systems subject to Coulomb frustration in the absence of 
global charge neutrality, from hot atomic nuclei to frustrated
ferromagnets~\cite{Bonche1985,Lee2001,Grousson,Raduta2002,Gulminelli2003}. 
	This discrepancy with the phase-transition 
phenomenology characteristic of atomic nuclei, indicates that a
connection between compact-star matter and nuclei is not trivial.
	Conversely a widening of the density range connected to the pasta 
phases was found~\cite{Maruyama2005} within the RMF model, when the 
Coulomb field is included under the constraint of global charge 
neutrality over the Wigner-Seitz cells.
	Such a finding is directly compatible with the 
thermodynamic phenomenology we are drawing.

	It was also discussed~\cite{Margueron2004,wata_rev} that neutrino propagation 
can be highly concerned by the presence of pasta phases and 
characterizes the process of neutron-star formation and cooling.
	With respect to homogeneous phases, the inhomogeneous 
phases (like pasta phases) present in the coexistence region are
characterized by a larger opalescence to neutrino propagation. 
	In the presence of a critical point 
	the opalescence would diverge as a direct consequence 
of the divergence of the correlation length, and neutrinos would be 
trapped.
	On the contrary, we conclude that the medium stays ``grey'' 
at any temperature.
	This result seems also connected to the observation of 
small opacity for the transport of neutrinos discussed in 
ref.~\cite{Horowitz2005}.
\end{document}